\documentclass[prb,showpacs,superscriptaddress,amsmath,amssymb,twocolumn]{revtex4-1}

\usepackage{graphicx}
\usepackage{epsfig}
\usepackage{subfigure}
\usepackage{hyperref}
\usepackage{epstopdf}
\usepackage{dcolumn}
\usepackage{bm}
\usepackage{color}

\begin{document}

\title{
\textit{Ab initio} calculation of valley splitting in monolayer $\delta$-doped phosphorus in silicon
}

\author{D.~W.~Drumm}
\email{d.drumm@student.unimelb.edu.au}
\affiliation{School of Physics, The University of Melbourne, Parkville 3010, Australia}

\author{A.~Budi}
\affiliation{School of Physics, The University of Melbourne, Parkville 3010, Australia}

\author{M.~C.~Per}
\affiliation{School of Applied Sciences, RMIT University, Melbourne 3001, Australia}

\author{S.~P.~Russo}
\affiliation{School of Applied Sciences, RMIT University, Melbourne 3001, Australia}

\author{L.~C.~L.~Hollenberg}
\affiliation{School of Physics, The University of Melbourne, Parkville 3010, Australia}



\date{\today}

\begin{abstract}

The differences in energy between electronic bands due to valley splitting are of paramount importance in interpreting transport spectroscopy experiments on state-of-the-art quantum devices defined by scanning tunneling microscope lithography.  We develop a plane-wave density functional theory description of these systems which is size-limited due to computational tractability.  We then develop a less resource-intensive alternative via localized basis functions, retaining the physics of the plane-wave description, and extend this model beyond the capability of plane-wave methods to determine the \textit{ab initio} valley splitting of well-isolated $\delta$-layers.  In obtaining agreement between plane-wave and delocalized methods, we show that the valley splitting has been overestimated in previous \textit{ab initio} calculations by more than 50\%.

\end{abstract}

\maketitle
\section{Introduction}
\label{sec:intro}

Quantum devices in silicon have been the subject of concentrated interest, both experimental and theoretical, in recent years.  Efforts to make such devices have led to atomically precise fabrication methods which incorporate phosphorus atoms in a single monolayer of a silicon crystal \cite{Tucker98,OBrien01,Shen02,Fuechsle07}.  These dopant atoms can be arranged into arrays \cite{Pok07}, or geometric patterns for wires \cite{Ruess07} and associated tunnel junctions \cite{Ruess07a}, gates, and quantum dots \cite{Ruess07c,Fuhrer09} -- all of which are necessary components of a functioning device \cite{Fuechsle10}.  The patterns themselves define atomically abrupt regions of doped and undoped silicon.  While silicon, bulk-doped silicon, and the physics of the phosphorus incorporation \cite{Wilson04} are well-understood, models of this quasi-two-dimensional phosphorus sheet are still in their initial stages.  In particular, it is critical in many applications to understand the effect of this confinement on the conduction band valley degeneracy, inherent in the band structure of silicon.  For example, the degeneracy of the valleys has the potential to cause decoherence in a spin-based quantum computer~\cite{Koiller01,Boykin04}, and the degree of valley degeneracy lifting (valley splitting) defines the conduction properties of highly confined planar quantum dots\cite{Fuechsle10}. 

The importance of understanding valley splitting in monolayer $\delta$-doped Si:P structures has led to a number of theoretical works in recent years spanning several techniques, from pseudopotential theories via planar Wannier orbital (PWO) bases \cite{Qian05}, density functional theory (DFT) via linear combination of atomic orbital (LCAO) bases \cite{Carter09,Carter11}, to tight-binding (TB) models \cite{Cartoixa05,Lee09,Ryu10,Ryu11,Lee11}, and effective mass theories (EMT) \cite{Scolfaro94,Rodriguez06,Drumm11}.  We note that several of these papers are based upon the assumption that the effective masses of $\delta$-doped P in Si remain unchanged from bulk-doped values;\cite{Scolfaro94,Rodriguez06} an assumption which has been challenged.\cite{Qian05,Cartoixa05}  Others assume doping over a multi-atomic plane band\cite{Cartoixa05,Scolfaro94} which no longer represents the state of the art in fabrication.  We also note that Ref. \onlinecite{Carter09} represents the first attempt to model these devices by considering explicitly doped $\delta$-layers with DFT, using a relatively small localized basis set with the assumption that a basis set sufficient to describe bulk silicon would also adequately describe P-doped Si.  There is currently little agreement between the valley splitting values obtained using these methods, with predictions ranging between 5 to 270 meV, depending on the arrangement of dopant atoms within the $\delta$-layer.  Density functional theory has been shown to be a useful tool in predicting how quantum confinement and/or doping perturbs the bulk electronic structure in silicon- and diamond-like structures,\cite{Delley93,Delley95,Ramos01,Zhou03,Barnard03c} and it might be expected that the removal of the basis set assumption will lead to the best estimate of the valley splitting available.

In this paper we determine a consistent value of the valley splitting in explicitly $\delta$-doped structures by obtaining convergence between distinct DFT approaches in terms of basis set and system sizes.  We perform a comparison of DFT techniques, involving localized numerical atomic orbitals and delocalized plane-wave (PW) basis sets.  Convergence of results with regard to the amount of Si ``cladding'' about the $\delta$-doped plane is studied.  This corresponds to the normal criterion of supercell size, where periodic boundary conditions may introduce artificial interactions between replicated dopants in neighboring cells.  A benchmark is set via the delocalized basis for DFT models of $\delta$-doped Si:P against which the localized basis techniques are assessed.  Implications for the type of modeling being undertaken are discussed, and the models extended beyond those tractable with plane-wave techniques.  Using these calculations, we obtain converged values for properties such as bandstructures, energy levels, valley splitting, electronic densities of state and charge densities near the $\delta$-doped layer. 

The paper is organized as follows: Sec. \ref{sec:methods} outlines the parameters used in our particular calculations; we present the results of our calculations in Sec. \ref{sec:blah}; and conclusions are drawn in Sec. \ref{sec:conclusion}.  An elucidation of effects modifying the bulk bandstructure follows in App. \ref{sec:bravais} \& \ref{sec:zfold} to provide a clear contrast to the properties deriving from the $\delta$-doping of the silicon discussed in the paper.

\section{Methodology}
\label{sec:methods}

Density functional theory calculations have been carried out using both plane-wave and LCAO basis sets.  For the plane-wave (PW) basis set, the Vienna \textit{ab initio} simulation package (\textsc{vasp}) \cite{Kresse99} software was used with projector augmented wave (PAW) \cite{Blochl94a,Kresse99} pseudopotentials for Si and P.  Due to the nature of the plane-wave (PW) basis set, there exists a simple relationship between the cutoff energy and basis set completeness.  For the structures considered in this work, the calculations were found to be converged for PW cutoffs of 450 eV.  

Localized basis set calculations were performed using the Spanish Initiative for Electronic Simulations with Thousands of Atoms (\textsc{siesta}) \cite{Artacho08} software.  In this case, the P and Si ionic cores were represented by norm-conserving Troullier-Martins pseudopotentials.\cite{Troullier91}  The Kohn-Sham orbitals were expanded in the default single-$\zeta$ polarized (SZP) or double-$\zeta$ polarized (DZP) basis sets, which consist of 9 and 13 basis functions per atom respectively.  Both the SZP and DZP sets contain $s$-, $p$-, and $d$-type functions.  These calculations were found to be converged for a mesh grid energy cutoff of 300 Ry.  In all cases, the generalized gradient approximation (GGA) PBE \cite{Perdew96} exchange-correlation functional was used.

The lattice parameter for bulk Si was calculated using an 8-atom cell, and found to be converged for all methods with a $12 \times 12 \times 12$ Monkhorst--Pack (MP) \textit{k}-point mesh \cite{Monkhorst76}.  The resulting values are presented in Table \ref{tab:lattice}, and were used in all subsequent calculations.

\begin{table}[ht]
  \begin{center}
    \begin{tabular}{cc}
      \hline
      \hline
      Method               & $a_{0}$ (\r{A})\\
      \hline
      PW (\textsc{vasp})   & 5.469 \\
      DZP (\textsc{siesta})& 5.495 \\
      SZP (\textsc{siesta})& 5.580 \\
      \hline
      \hline
    \end{tabular}
  \end{center}
  \caption{Equilibrium lattice parameters for an 8-atom cubic unit cell for the different methods used in this work.}
  \label{tab:lattice}
\end{table}

\begin{figure}[ht]
\includegraphics[width=0.6\linewidth]{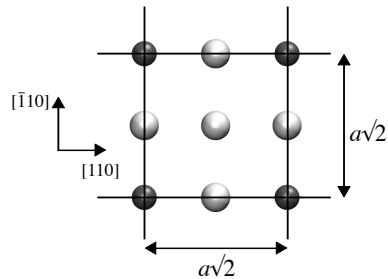}
\caption{$\left(001\right)$ planar slice of the $c\left(2\times2\right)$ structure, at the 1/4 ML doped monolayer.  One of the Si sites has been replaced by a P atom (shown in dark gray). The periodic boundaries are shown in black.}
\label{fig:c22schematic}
\end{figure}

In modeling $\delta$-doped Si:P, as used in Ref. \onlinecite{Fuechsle10}, we adopted a tetragonal supercell description of the system, akin to that of Refs. \onlinecite{Qian05} \& \onlinecite{Carter09}.  In accordance with experiment, we inserted the P layer in a monatomic $\left(001\right)$ plane as one atom in four to achieve 25\% doping.  This will henceforth be referred to as 1/4 monolayer (ML) doping.  In this case, the smallest repeating in-plane unit had four atoms per monolayer (to achieve 1 in 4 doping), and was a square with sides parallel to the $\left[110\right]$ and $\left[\bar{1}10\right]$ directions.  The square had side length $a\sqrt{2}$ (see Fig. \ref{fig:c22schematic}), where $a$ is the simple cubic lattice constant of bulk silicon.  The phosphorus layers had to be separated by a considerable amount of silicon due to the large Bohr radius of the hydrogen-like orbital introduced by P in Si ($\sim$2.5 nm).  Ref. \onlinecite{Carter09} showed that this far exceeded the sub-nanometer cell side length.  If desired, cells with a lower in-plane density of dopants may be constructed, by lengthening the cell in the $x$- and $y$-directions, such that more Si atoms occupy the doped monolayer in the cell.


%

\begin{figure}[ht]
\includegraphics[width=0.6\linewidth]{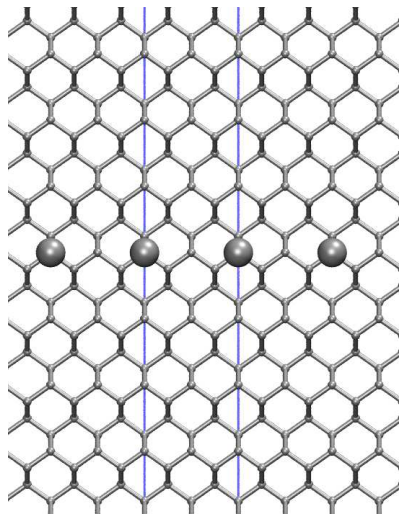}
\caption{Ball \& stick model of a $\delta$-doped Si:P layer, viewed along the $[110]$ direction; 32 layers in the $[001]$ direction are shown.  Si atoms (small gray spheres), P atoms (large dark gray spheres), covalent bonds (gray sticks), repeating cell boundary (solid line).}
\label{fig:diagram_P_layer_pbc}
\end{figure}

A collection of tetragonal cells comprised of 4, 8, 16, 32, 40, 60, 80, 120, 160 and 200 monolayers were constructed, having four atomic sites per monolayer and oriented with faces in the $[110]$, $[\bar{1}10]$ and $[001]$ directions (see Fig. \ref{fig:diagram_P_layer_pbc}).  Cells used in PW calculations began at 4 layers and ran to 80 layers; larger cells were not computationally tractable with this method.  SZP and DZP models began at 40 layers to overlap with PW for the converging region, and were then extended to their tractable limit (200 and 160 layers, respectively) to study convergence past the capability of PW.

For the tetragonal cells the \textit{k}-point sampling was set as a $9\times9\times N$ $\Gamma$-centred MP mesh, as we have found that failing to include $\Gamma$ in the mesh can lead to anomalous placement of the Fermi level on bandstructure diagrams.  $N$ varied from 12 to 1 as the cells became more elongated (see Table \ref{tab:bulkSi} in App. \ref{sec:bravais}).

Although it has been previously found that relaxing the positions of the nuclei gave negligible differences ($<$~0.005 \r{A}) to the geometry,\cite{Carter09} this was for a 12-layer cell and may not have included enough space between periodic repetitions of the doping plane for the full effect to be seen.  We have performed a test relaxation on a 40-layer cell using the PW basis (\textsc{vasp}).  The maximum subsequent ionic displacement was 0.05 \r{A}, with most being an order of magnitude smaller.  The energy gained in relaxing the cell was less than 37 meV (or 230 $\mu$eV/atom).  We therefore regarded any changes to the structure as negligibly small, and proceeded without ionic relaxation.

Single-point energy calculations were carried out with both software programs; for \textsc{vasp} the electronic energy convergence criterion was set to $10^{-6}$ eV, and the tetrahedron method with Bl\"ochl correction \cite{Blochl94} was used.  For \textsc{siesta} a two-stage process was carried out: Fermi-Dirac electronic smearing of 300 K was applied in order to converge the density matrix within a tolerance of 1 part in $10^{-4}$; the calculation was then restarted with smearing of 0 K and a new electronic energy tolerance criterion of $10^{-6}$ eV applied (except for the 120- and 160-layer DZP models for which this was intractable; a tolerance of $10^{-4}$ eV was used in these cases).  This two-stage process aided convergence as well as ensuring that the energy levels obtained were comparably accurate across methods.  In addition, for each doped cell thus developed and studied, an undoped bulk Si cell of the same dimensions was constructed to aid in isolating those features primarily due to the doping.

\section{Results}
\label{sec:blah}

\subsection{Analysis of bandstructure}
\label{sec:results}

Once converged charge densities were obtained, bandstructures were calculated along the $M$--$\Gamma$--$X$ high-symmetry pathway (as shown in Fig. \ref{fig:tetBZ} in Appendix \ref{sec:bravais}), using at least 20 \textit{k}-points between high-symmetry points.  For comparative purposes, the bandstructures have all been aligned at the valence band maximum (VBM).

\begin{figure}[th]
\includegraphics[angle=270, width=\linewidth]{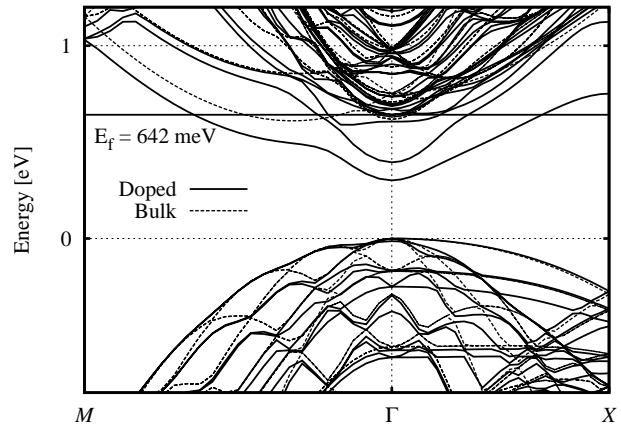}
\label{fig:d1vasp}
\caption{Full band structure of the 40-layer tetragonal system, calculated using PW (\textsc{vasp}).  Bulk and 1/4 ML doped structures are shown.}
\label{fig:bandscomparison}
\end{figure}

Figure \ref{fig:bandscomparison} contrasts bulk and doped bandstructures for the 40-layer PW calculation.  DZP and SZP results are similar on this scale and are omitted in the interest of clarity in the diagram.  As discussed in App. \ref{sec:zfold}, it is evident from the bulk values that the elongated cells have led to the folding of two CBM valleys towards the $\Gamma$-point.  Also visible is the difference the doping potential makes to the system; what was the lowest unoccupied orbital in the bulk is now dragged down in energy by the extra ionic potential.  It is of note that the region near $\Gamma$, corresponding to the $k_{z}$ valleys which can be modeled as having different effective masses to the $k_{x,y}$ valleys,\cite{Qian05} is brought lower than the region corresponding to the $k_{x,y}$ valleys and is non-degenerate.  The second band behaves in a similar fashion.  The third band appears to maintain a minimum away from the $\Gamma$-point in the $\Sigma_{\rm{TET}}$-direction (which is equivalent to the $\Delta_{\rm{FCC}}$-direction; see App. \ref{sec:bravais}), but in a less-parabolic fashion than the lower two; its minimum is similar to the value at $\Gamma$.  This band is non-degenerate along this particular direction in $k$-space, but due to the supercell symmetry it is actually 4-fold degenerate, in contrast to the other bands.  
The Fermi level for the doped system is also shown, clearly being crossed by all three of these bands which are therefore able to act as open channels for conduction.

\begin{figure}[bh]
\includegraphics[angle=270, width=\linewidth]{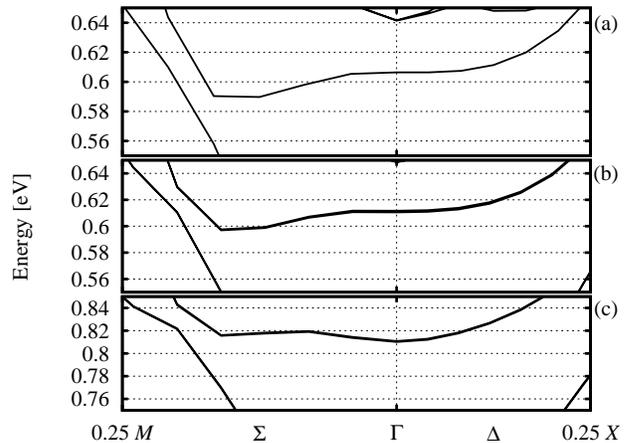}
\caption{Bandstructure of the 40-layer tetragonal system, zoomed in on the $\Delta$ band, calculated with (a) PW (\textsc{vasp}), (b) DZP (\textsc{siesta}), and (c) SZP basis sets.}
\label{fig:zoom}
\end{figure}

As mentioned above, the bandstructures are similar across all methods, but upon detailed inspection important differences come to light.  A closer look at the $\Delta$ band shows a qualitative difference between the predictions using SZP (Fig. \ref{fig:zoom}c) and the PW and DZP results (Figs. \ref{fig:zoom}a \& \ref{fig:zoom}b): the models with a more complete basis predict the band minimum to occur in the $\Sigma_{\rm{TET}}$ ($\Delta_{\rm{FCC}}$) direction, below the value at $\Gamma$, while the SZP bandstructure shows the reverse -- the minimum at $\Gamma$, a similar amount below a secondary minimum in the $\Sigma_{\rm{TET}}$ direction.  

\begin{figure}[bh]
\mbox{\subfigure[]{\label{fig:dopedPAW}\includegraphics[width=0.8\linewidth]{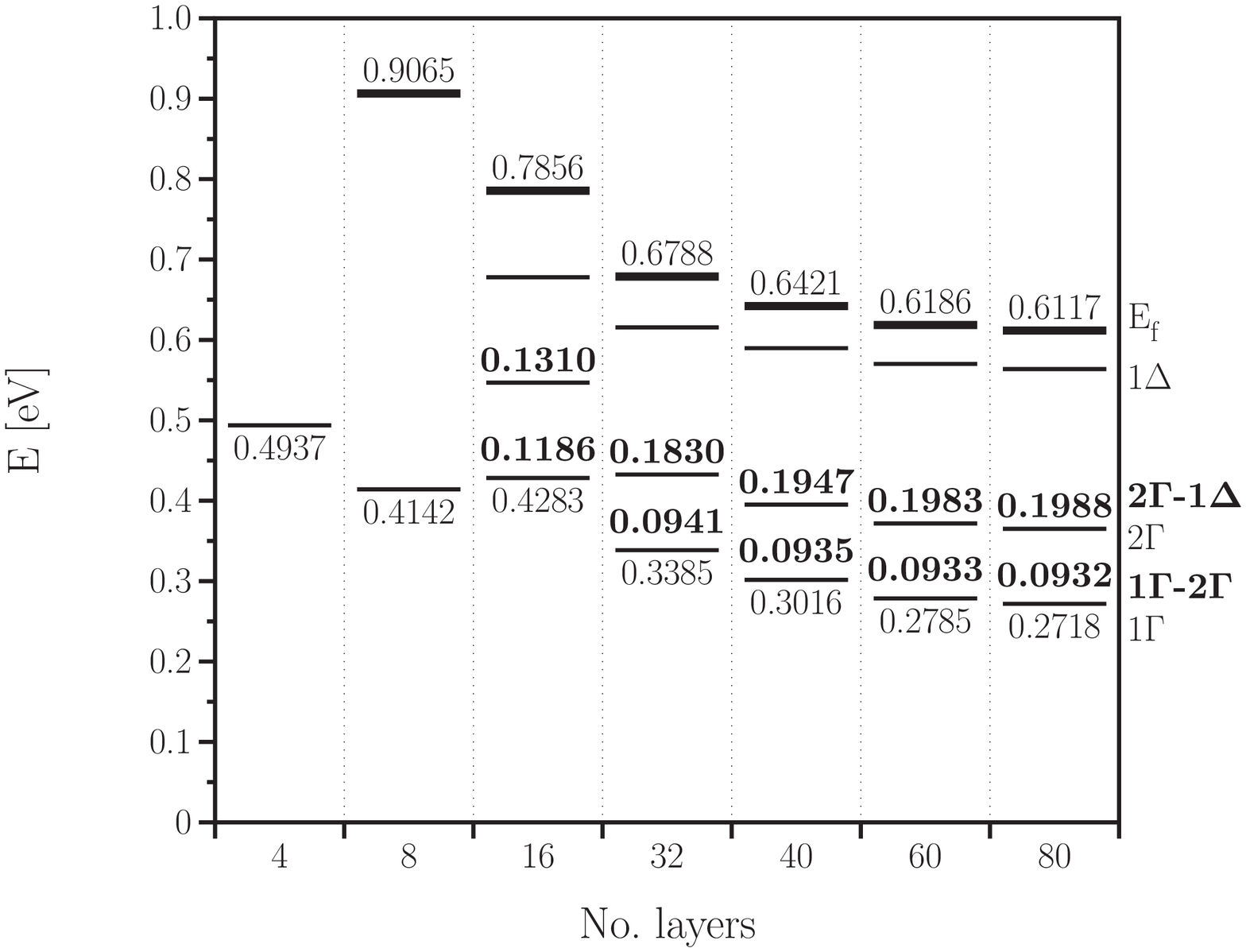}}}
\mbox{\subfigure[]{\label{fig:dopedDZP}\includegraphics[width=0.8\linewidth]{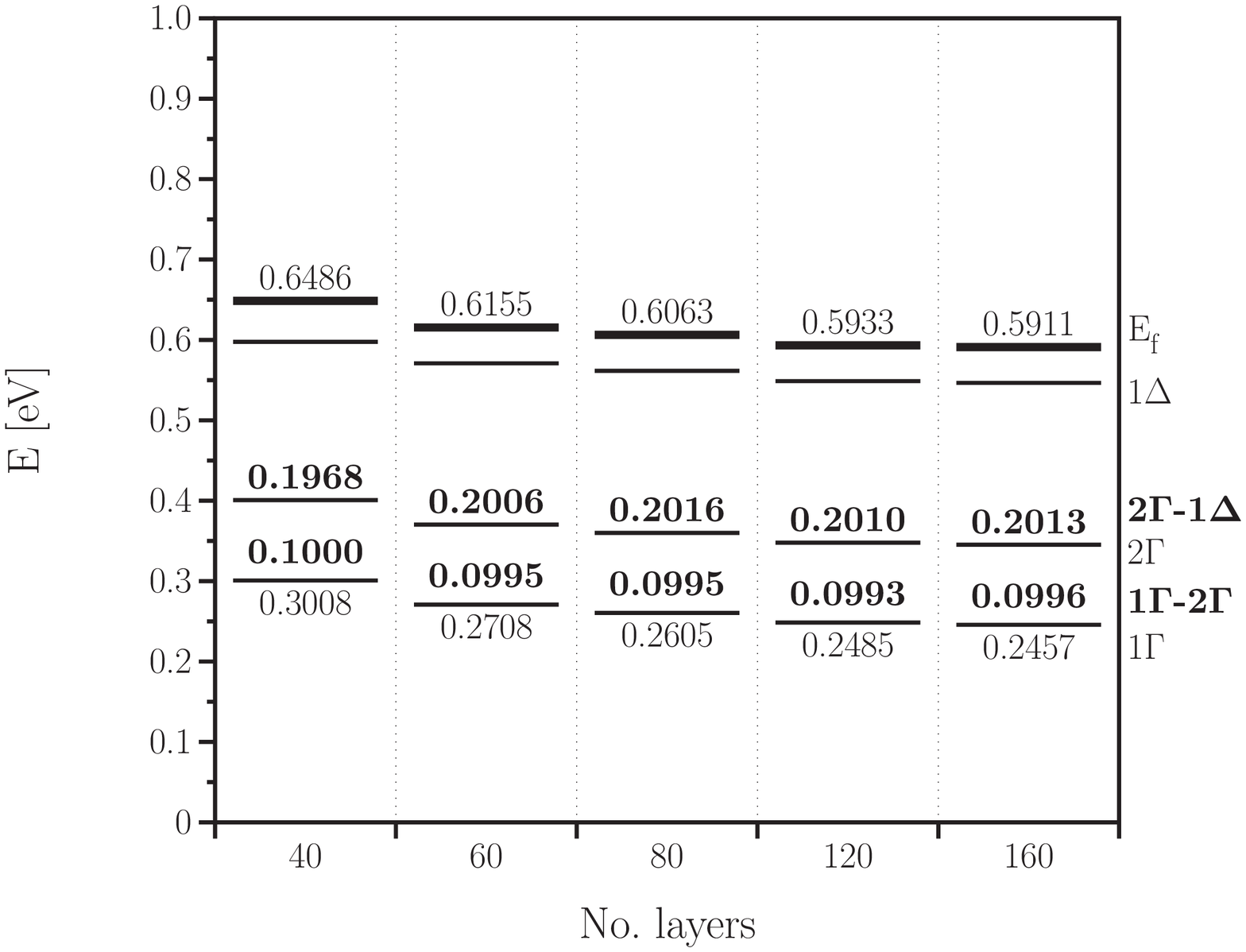}}}
\mbox{\subfigure[]{\label{fig:dopedSZP}\includegraphics[width=0.8\linewidth]{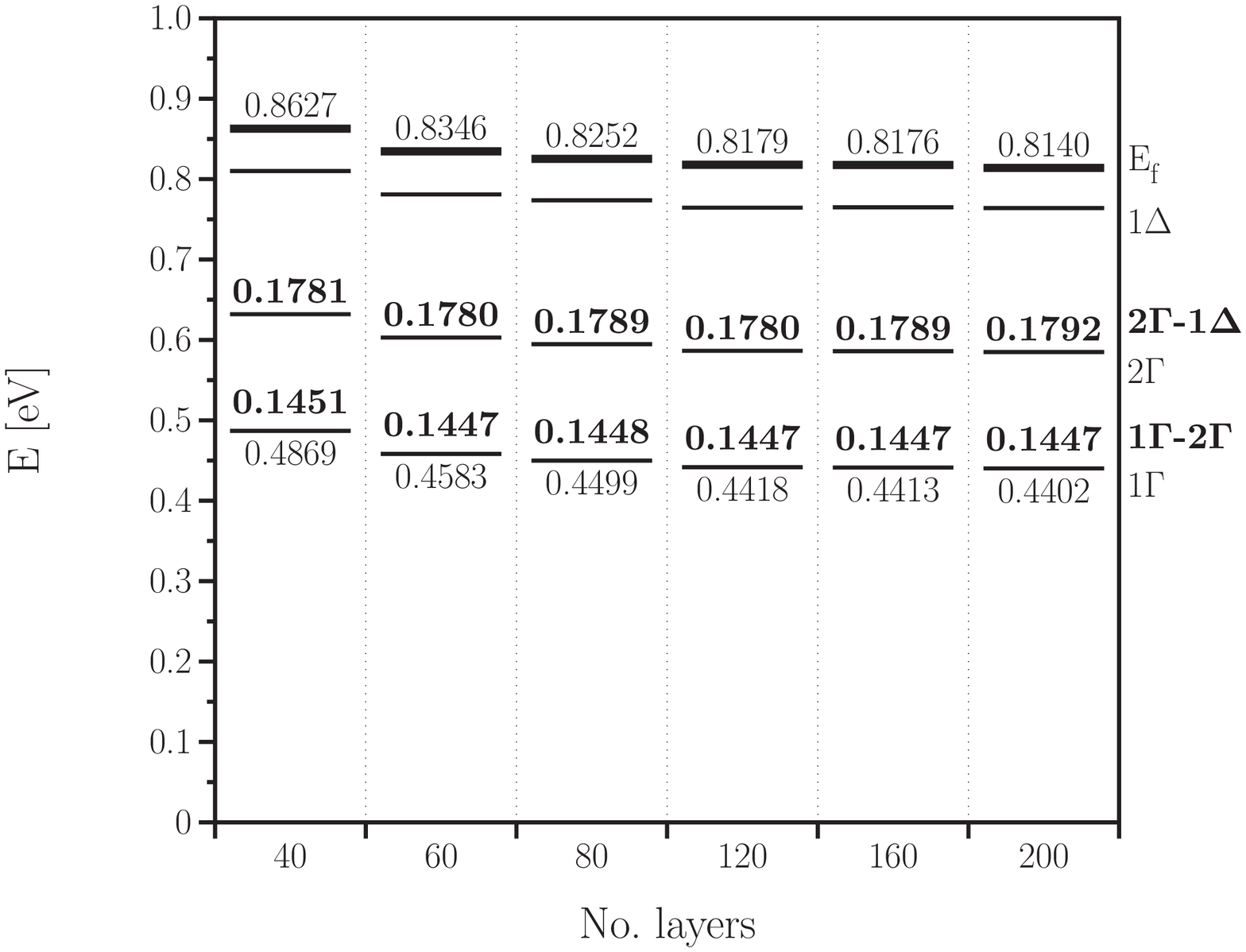}}}
\caption{Minimum band energies for tetragonal systems with 1/4 ML doping, calculated using : (a) PW (\textsc{vasp}), (b) DZP (\textsc{siesta}), and (c) SZP (\textsc{siesta}) basis sets.  Fermi level also shown where appropriate.  Bold numbers indicate energy differences between band minima.}
\label{fig:energies}
\end{figure}

The difference between the energies of the first two band minima ($\Gamma_{1}$--$\Gamma_{2}$, illustrated in Fig. \ref{fig:energies}), or the valley splitting, from the PW and DZP calculations agree with each other to within $\sim$6 meV. Significantly, the value obtained using the SZP basis set differs by 52 meV, some 55\% larger than the value obtained using the PW basis set.  The importance of this discrepancy cannot be overstated; this valley splitting is directly relatable to experimentally observable resonances in transport spectroscopy of devices made with this $\delta$-doping technology (see Ref. \onlinecite{Fuechsle10}).

In the smallest cells ($<$ 16 layers), less than three bands are observed.  This is likely due to the lack of cladding in the $z$-direction, leading to significant interaction between the dopant layers, raising the energy of each band.  Whilst the absolute energy of each level still varies somewhat, even with over 100 layers incorporated, we find that the $\Gamma_{1}$--$\Gamma_{2}$ values are well-converged with 80 layers of cladding for all methods (see Fig. \ref{fig:energies}).  Indeed, they may be considered reasonably converged even at the 40-layer level (0.5 meV or less difference to the largest models considered).  The differences between the energies of the second and third band minima ($\Gamma_{2}$--$\Delta$ splittings) are also shown in Fig. \ref{fig:energies}, and show good convergence (within 1 meV) for cells of 80 layers or larger.

The Fermi level follows a similar pattern to the $\Gamma$- and $\Delta$-levels.  In particular, the gap between the Fermi level and $\Gamma_{1}$ level does not change by more than 1 meV from 60 to 160 layers.

Given that the properties of interest are the differences between the energy levels, rather than their absolute values (or position relative to the valence band), in the interest of computational efficiency we observe that using the DZP basis with 80 layers of cladding is sufficient to achieve consistent, converged results.

\subsection{Valley splitting}
\label{sec:valley}

Table~\ref{tab:valley} summarizes the valley splitting values of 1/4 ML P-doped silicon obtained using different techniques, showing a large variation in the actual values.  In order to make sense of these results, it is important to note two major factors that affect valley splitting: the doping method and the arrangement of phosphorus atoms in the $\delta$-layer.  As the results from Ref.~\onlinecite{Carter11} show, the use of implicit doping causes the valley splitting value to be much smaller than in an explicit case ($\sim$7 meV vs. 120 meV).  It is also shown that the use of random P coverage on the $\delta$-layer reduces the valley splitting value by only 40--50 meV compared to the fully ordered placement, leaving a large discrepancy between the valley splitting results from implicit and explicit doping.  This massive decrease in valley splitting due to implicit doping can be explained by the smearing of the doping layer in the direction  normal to the $\delta$-layer, thereby decreasing the quantum-confinement effect responsible for breaking the degeneracy in the system.  Ref.~\onlinecite{Carter11} also shows that the arrangement of the phosphorus atoms in the $\delta$-layer strongly influences the valley splitting value.  In particular, they showed that there is a difference of up to 220 meV between P doping along the $[110]$ direction and along the $[100]$ direction.  It must be noted, however, that most of their patterns are not yet physically realizable due to the P incorporation mechanism currently employed.

\begin{table}[tbh]
\renewcommand{\thefootnote}{\thempfootnote}
	\begin{center}
		\begin{tabular}{ccc}
		\hline
		\hline
		Technique & No. of 				& Valley  	\\
							& layers				&	splitting \\
							&								& (meV)			\\
		\hline
		Planar Wannier orbital\footnote{implicit doping}$^{,}$\cite{Qian05}	&  1000  &  20 \\
		\hline
		Tight binding (4 K)\footnote{explicit doping}$^{,}$\cite{Lee09}			&  $\sim$150  &  $\sim$17 \\
		Tight binding (4 K)$^{b,}$\cite{Lee11} 															&  120				&	25 \\
		\hline
		Tight binding (300 K)\footnotemark[\value{mpfootnote}]$^{,}$\cite{Ryu11}		&  $\sim$150  &  $\sim$17 \\
		\hline
								&    40  &   7 \\
								&    80  &   6 \\
		DFT, SZP basis set~$^{a,}$\cite{Carter11}&   120  &   6 \\
								&   160  &   6 \\
								&   200  &   6 \\
		\hline
		ordered~$^{b,}$\cite{Carter09}~~~~~ & 40 & 120 \\
		~~~~~~~random disorder~$^{b,}$\cite{Carter09} &    40  &  $\sim$70 \\
		~~~~~~~~~~~~~~~~~~$[110]$ direction alignment~$^{b,}$\cite{Carter11}	&    40  & $\sim$270 \\
		DFT, SZP: dimers~$^{b,}$\cite{Carter11}~~~~~~~~~~~~~~~~~~~~~		&    40  &  $\sim$85 \\
		~~~~~~~random disorder~$^{b,}$\cite{Carter11}		&    40  &  $\sim$80 \\
		clusters~$^{b,}$\cite{Carter11}~~~~~		&    40  &  $\sim$65 \\
		~~~~~~~~~~~~~~~~~~$[100]$ direction alignment~$^{b,}$\cite{Carter11}	&    40  &  $\sim$50 \\
		\hline
		~~~~~ordered, $M$=4$^{b,}$\footnote{$M\times M\times 1$ $k$-points used}$^{,}$\cite{Carter11} & 80	&	153 \\	
 		DFT, SZP: ordered, $M$=6$^{b,}$\footnotemark[\value{mpfootnote}]$^{,}$\cite{Carter11}~~~~~~~~~~ & 80 & 147 \\
		~~~~~~~ordered, $M$=10$^{b,}$\footnotemark[\value{mpfootnote}]$^{,}$\cite{Carter11} & 80	& 147 \\
		\hline
								&    40  & 145.1 \\
								&    60  & 144.7 \\
		SZP, $M$=9 (this work)$^{b,c,}$						&    80  & 144.8 \\
								&   120  & 144.7 \\
								&   160  & 144.7 \\
								&   200  & 144.7 \\
		\hline
								&    16  & 118.6 \\
								&    32  &  94.1 \\
		PW, $M$=9 (this work)$^{b,}$\footnote{$M\times M\times N$ $k$-points used; $N$ as per Table \ref{tab:bulkSi} in App. \ref{sec:bravais}}	&    40  &  93.5 \\
								&    60  &  93.3 \\
								&    80  &  93.2 \\
		\hline
								&    40  & 100   \\
								&    60  &  99.5 \\
		DZP, $M$=9 (this work)$^{b,c}$	&    80  &  99.5 \\
								&   120  &  99.3 \\
								&   160  &  99.6 \\
		\hline
		\hline
		\end{tabular}
	\end{center}
	\caption{Valley splitting values of 1/4 ML P-doped silicon obtained using different techniques.  Techniques are grouped by similarity.}
	\label{tab:valley}
\end{table}

Our results show that valley splitting is highly sensitive to the choice of basis set. Due to the nature of PW basis set, it is straightforward to improve its completeness by increasing the plane wave cutoff energy.  In this way, we establish the most accurate valley splitting value within the context of density functional theory.  Using this benchmark value, we can then establish the validity and accuracy of other basis sets, which can be used to extend the system sizes to that beyond what is practical using PW basis set.  As seen in Table~\ref{tab:valley}, the valley splitting value converges to 93 meV using 80-layer cladding.  The DZP localized basis set gives an excellent agreement at 99.5 meV using 80-layer cladding (representing a 7\% difference).  On the other hand, the SZP localized basis set (similar to what was used in Refs.~\onlinecite{Carter09} and~\onlinecite{Carter11}) gave a value of 145 meV using the same amount of cladding.  This represents a significant difference of 55\% over the value obtained using PW basis set, and demonstrates that the SZP basis set is unsuitable for accurate determination of valley splitting in these systems.

\subsection{Density of states}
\label{sec:eDOS}

\begin{figure}[ht]
\includegraphics[angle=270,width=\linewidth]{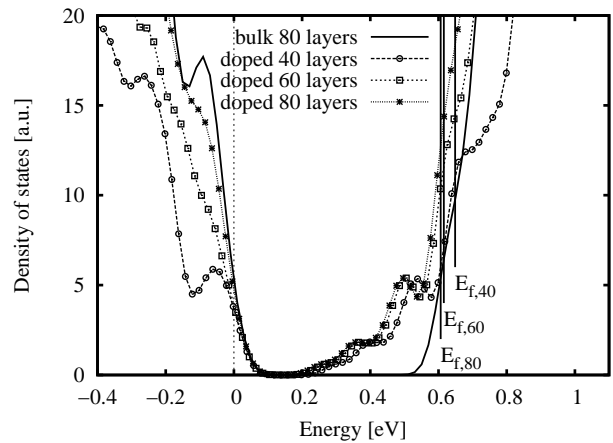}
\caption{Electronic densities of states for tetragonal systems with 0 and 1/4 ML doping, calculated using the DZP (\textsc{siesta}) basis set.  The Fermi level is indicated by a solid vertical line with label.  50 meV smearing was applied for visualization purposes.}
\label{fig:eDOS}
\end{figure}

The electronic density of states (eDOS) was calculated for each cell.  Figure \ref{fig:eDOS} compares the unscaled eDOS for bulk 80-layer cells to that of doped cells varying from 40 to 80 layers.  The bulk bandgap is visible, with the conduction band rising sharply to the right of the figure.  The doped eDOS exhibits density in the bulk bandgap, although the features of the spectra differ slightly according to the basis set used.

The Fermi energy exhibits convergence with respect to the amount of cladding, as reported above.  It is also notable that the eDOS within the bandgap are nearly identical regardless of the cell length (in $z$).  This indicates that layer-layer interactions are negligibly affecting the occupied states, and therefore that the applied ``cladding'' is sufficient to insulate against these effects.
%

%
%
\subsection{Electronic width of the plane}
\label{sec:width}

\begin{figure}[ht]
  \includegraphics[angle=270, width=0.8\linewidth]{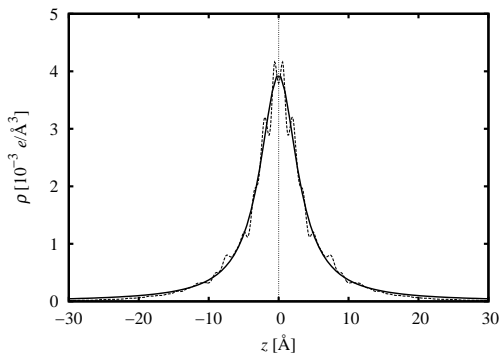}
  \caption{Planar average of the donor-electron density as a function of $z$-position for the 1/4 ML doped, 80-layer cell calculated using the DZP basis set.  The fitted Lorentzian function is also shown.}
  \label{fig:d1dzprho}
\end{figure}

In order to quantify the extent of the donor-electron distribution, we have integrated the local density of states (LDOS) between the VBM and Fermi level and taken the planar average with respect to the $z$-position.  Figure~\ref{fig:d1dzprho} shows the planar average of the donor electrons (a sum of both spin-up and spin-down channels) for the 80-layer cell calculated using the DZP basis set.  After removing the small oscillations related to the crystal lattice to focus on the physics of the $\delta$-layer, by Fourier transforming, a Lorentzian function was fitted to the distribution profile.  (Initially, a three-parameter Gaussian fit 
similar to that used in Ref. \onlinecite{Drumm11} was tested, but the Lorentzian gave a better fit to the curve.)

Table~\ref{tab:ldosrho} summarizes the maximum donor-electron density and the full width at half maximum (FWHM) for the 1/4 ML doped cells, each calculated from the Lorentzian fit.  Both of these properties are remarkably consistent with respect to the number of layers, indicating that they have converged sufficiently even at 40 layers. 

\begin{table}[ht]
  \begin{center}
    \begin{tabular}{ccc}
      \hline
      \hline
      No. of  & $\rho_{\textrm{max}}$   & FWHM      \\
      layers  & ($\times 10^{-3}$ $e$/\r{A}) & (\r{A})   \\
      \hline
       40      & 3.8         & 6.2       \\
       60      & 3.9         & 6.2       \\
       80      & 3.9         & 6.5       \\
      \hline
      \hline
    \end{tabular}
  \end{center}
  \caption{Calculated maximum donor-electron density, $\rho_{\textrm{max}}$, and full width at half maximum, FWHM, as a function of the number of layers in the 1/4 ML doped cells.  The DZP basis was used.}
  \label{tab:ldosrho}
\end{table}

Our results differ from a previous DFT calculation~\cite{Carter11} which cited a FWHM of 5.6 \r{A} for a 1/4 ML doped, 80-layer cell calculated using the SZP basis set (and 10$\times$10$\times$1 $k$-points).  We note that those values were taken from the unfitted, untransformed donor-electron distribution, and represent a $\sim$15\% underestimation of the DZP result.  

\section{Conclusion}
\label{sec:conclusion}


In this article, we have studied the valley splitting of monolayer $\delta$-doped Si:P, using a density functional theory model with a plane-wave basis to establish firm grounds for comparison with less computationally-intensive localized basis \textit{ab initio} methods.  We found that the best current descriptions of these systems (by density functional theory, using SZP basis functions) overestimate the valley splitting by over 50\%, due to an assumption made early in their methodology.  We show that DZP basis sets are complete enough to deliver values within 10\% of the plane-wave values, and due to their localized nature, are capable of calculating the properties of models twice as large as is tractable with plane-wave methods.  These DZP models are converged with respect to size well before their tractable limit, which approaches that of SZP models.

Valley splittings are important in interpreting transport spectroscopy experiment data, where they relate to families of resonances, and in benchmarking other theoretical techniques more capable of actual device modeling.  It is therefore pleasing to have an \textit{ab initio} description of this effect which is fully-converged with respect to basis completeness, as well as the usual size effects and $k$-point mesh density.

We have also studied the bandstructures with all three methods, finding that the DZP correctly determines the $\Delta$-band minima away from the $\Gamma$ point, where the SZP method does not.  We show that these minima occur in the $\Sigma$ direction for the type of cell considered, not the $\Delta$ direction as has been previously reported.  Having established the DZP methodology as sufficient to describe the physics of these systems, we then calculated the electronic density of states, and the electronic width of the $\delta$-layer.  We found that previous SZP descriptions of these layers underestimate the width of the layers by almost 15\%.

We have shown that the properties of interest of $\delta$-doped Si:P are well-converged for 40-layer supercells using a DZP description of the electronic density.  We recommend the use of this amount of surrounding silicon, and technique, in any future DFT studies of these and similar systems - especially if inter-layer interactions are to be minimized.

\section*{Acknowledgements}

The authors acknowledge funding by the ARC Discovery grant DP0881525. This research was undertaken on the NCI National Facility in Canberra, Australia, which is supported by the Australian Commonwealth Government.

\appendix

\section{Subtleties of bandstructure}
\label{sec:bravais}


Regardless of the type of calculation being undertaken, a bandstructure diagram is inherently linked to the type (shape and size) of cell being used to represent the system under consideration.  For each of the 14 Bravais lattices available for three-dimensional supercells, a particular Brillouin zone (BZ) with its own set of high-symmetry points exists in reciprocal space \cite{Bradley72}.  Similarly, each BZ has its own set of high-symmetry directions.  Some of these BZs share a few high-symmetry point labels (or directions), such as $X$ or $L$ ($\Delta$ or $\Sigma$), and they all contain $\Gamma$, but these points are not always located in the same place in reciprocal space.

A simple effect of this can be seen by increasing the size of a supercell.  This has the result of shrinking the BZ, and the coordinates of high-symmetry points on its boundary, by a corresponding factor.  Consider the conduction band minimum (CBM) found at the $\Delta$ valley in the Si conduction band.  This is commonly located at $k_{0}\sim0.85\frac{2\pi}{a}$ in the $\Delta$ direction towards $X$.  Should we increase the cell by a factor of 2, the BZ will shrink (BZ$\rightarrow$BZ$^{'}$), placing the valley outside the new BZ boundary (past $X^{'}$); but a valid solution in any BZ must be a solution in all BZs.  This results in the phenomenon of band folding, whereby a band continuing past a BZ boundary reenters the BZ on the opposite side.  Since the $X$ direction in a face-centred cubic (FCC) BZ is 6-fold symmetric, a solution near the opposite BZ boundary is also a solution near the one we are focussing on.  This results in the appearance that the band continuing past the BZ boundary is ``reflected'', or folded, back on itself into the first BZ.  Since the new BZ boundary in this direction is now at $k_{\rm{BZ}}^{'}=X^{'}=0.5\frac{2\pi}{a}$, the location of the valley will be at $k_{0}^{'}=X^{'}-\left(k_{0}-X^{'}\right)\sim0.15\frac{2\pi}{a}$, as mentioned in Ref. \onlinecite{Carter09}.  Each further increase in the size of the supercell will result in more folding (and a denser bandstructure).  Care is therefore required to distinguish between a new band and one which has been folded due to this effect when interpreting bandstructure.

\begin{figure}[b!]
\mbox{\subfigure[]{\label{fig:folding}\includegraphics[angle=270, width=\linewidth]{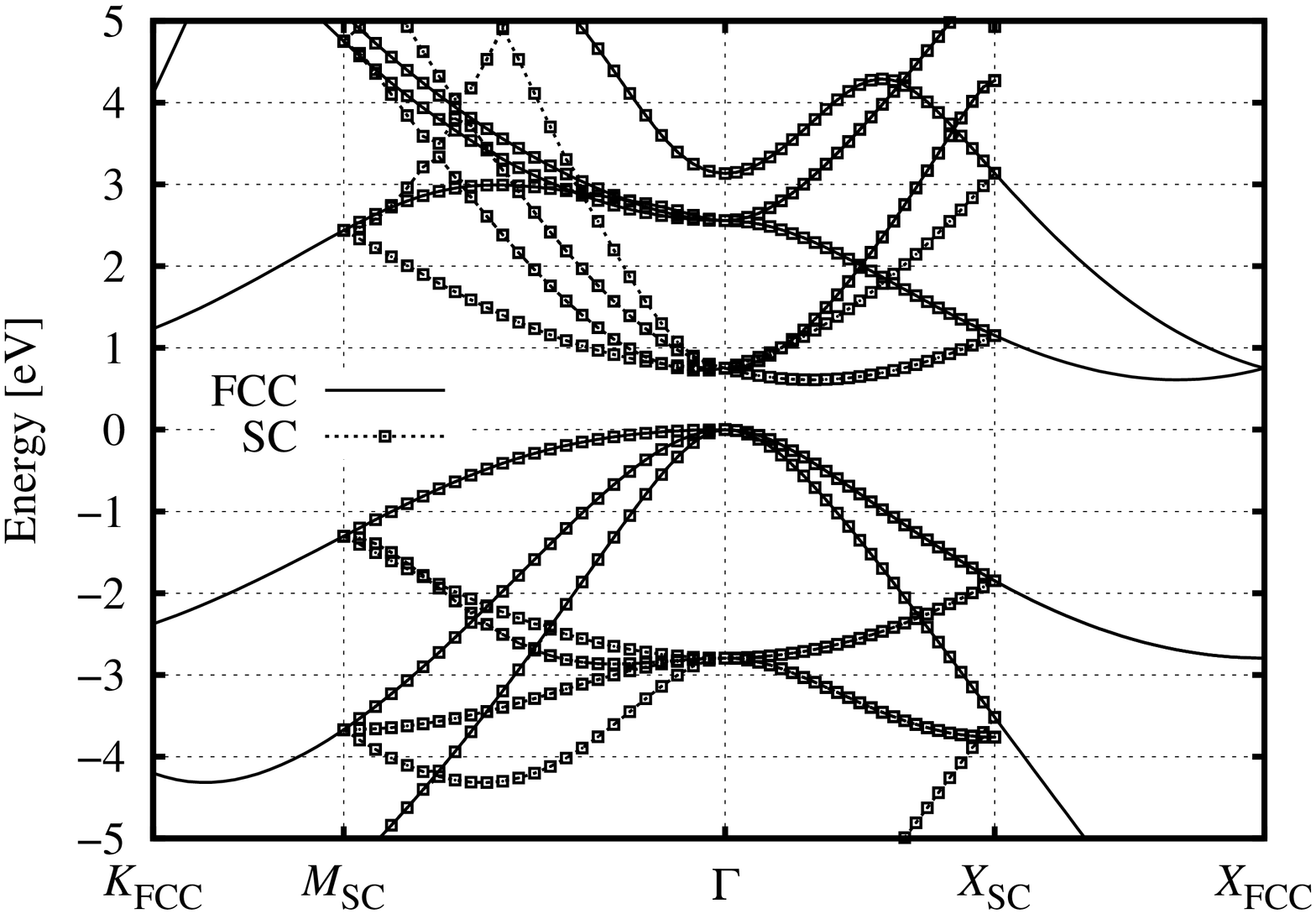}}}
\mbox{\subfigure[]{\label{fig:fcc}\includegraphics[angle=0, width=0.5\linewidth]{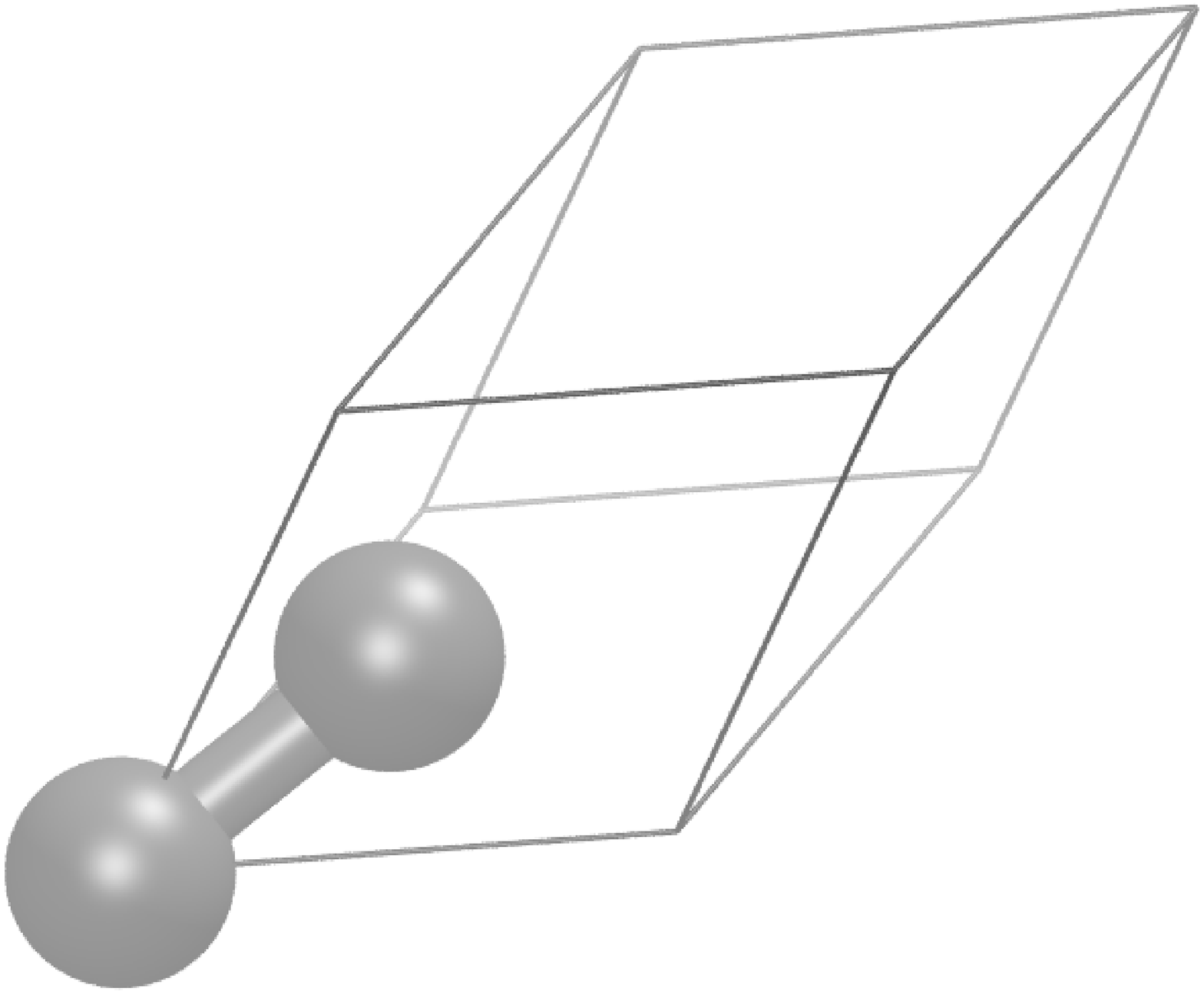}}\subfigure[]{\label{fig:p11}\includegraphics[angle=0, width=0.5\linewidth]{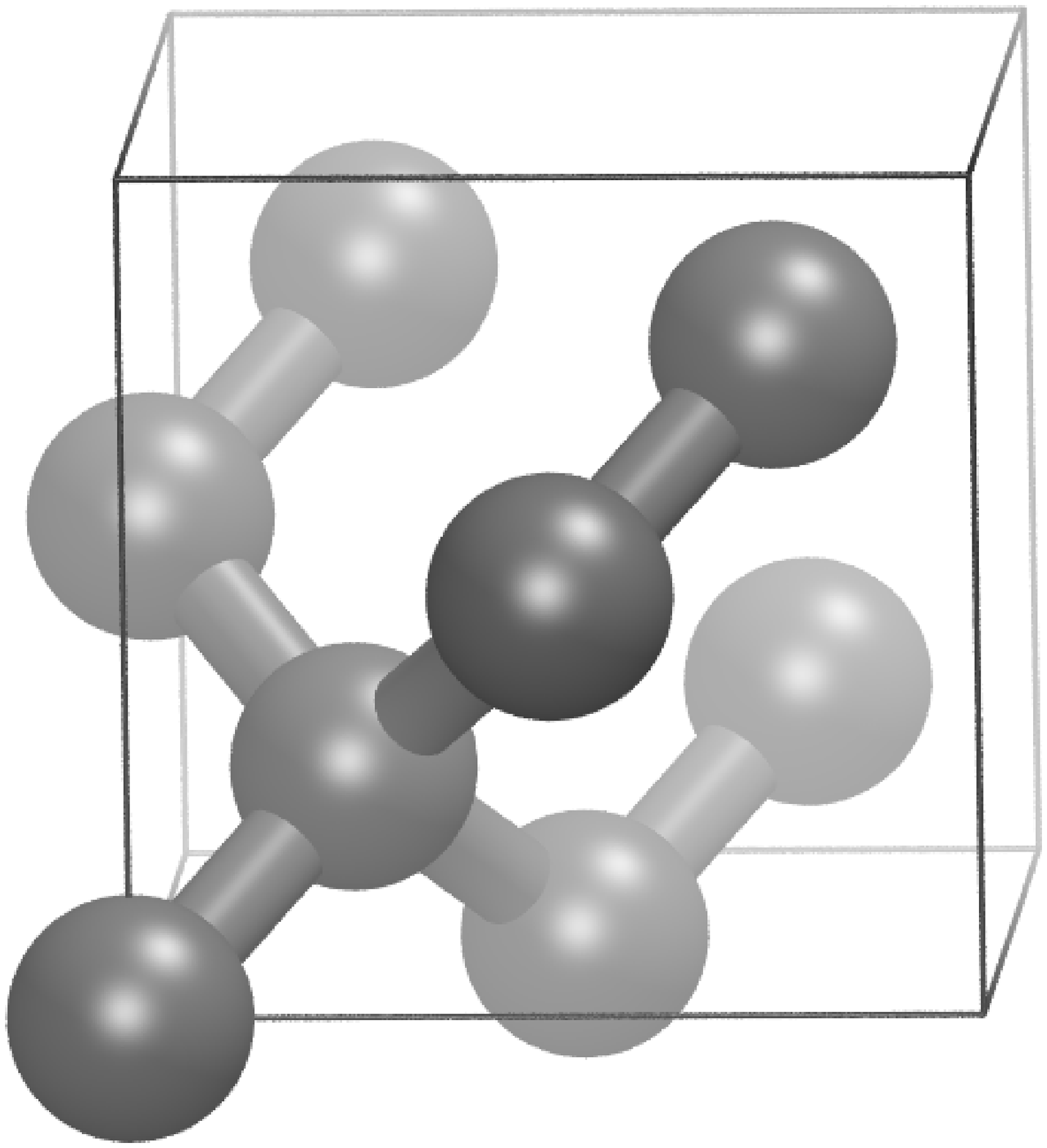}}}
\caption{(a) Typical band structure of bulk Si for 2-atom FCC (solid lines) and 8-atom SC cells (dotted lines with squares), calculated using the \textsc{vasp} plane-wave method (see Sec. \ref{sec:methods}).  (b) 2-atom FCC cell.  (c) 8-atom SC cell.}
\label{fig:bandstructure}
\end{figure}

Continuing with our example of silicon, whilst the classic bandstructure \cite{Chelikowsky74} is derived from the bulk Si primitive FCC cell (containing two atoms), it is often more convenient to use a simple cubic (SC) supercell (8 atoms) aligned with the $\langle1 0 0\rangle$ crystallographic directions.  In this case, we experience some of the common labelling; the $\Delta$ direction is defined in the same manner for both BZs, although we see band folding (in a similar manner to that discussed above) due to the size difference of the reciprocal cells (see Fig. \ref{fig:bandstructure}).  We also see a difference in that although the $\Sigma$ direction is consistent, the points at the BZ boundaries have different symmetries and therefore, labels ($K_{\rm{FCC}}$, $M_{\rm{SC}}$).  (The $L_{\rm{FCC}}$-point and $\Lambda_{\rm{FCC}}$-direction have no equivalent for tetragonal cells, and hence we do not consider bandstructure in that direction here)

\begin{figure}[t!]
\includegraphics[width=0.5\linewidth]{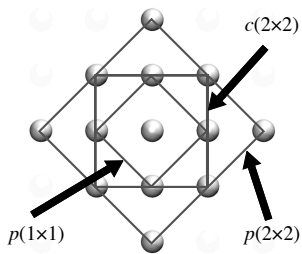}
\caption{Geometrical difference between the simple cubic and tetragonal cells; $\left(0 0 1\right)$ planar cut through an atomic monolayer.}
\label{fig:rotation}
\end{figure}

\begin{figure}[h]
\includegraphics[width=0.6\linewidth]{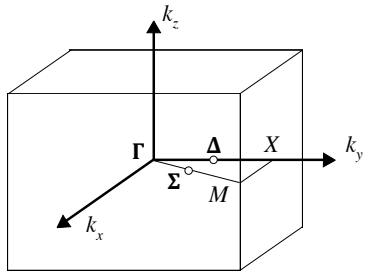}
\caption{The Brillouin zone for a tetragonal cell. The $M$--$\Gamma$--$X$ path used in this work is shown.}
\label{fig:tetBZ}
\end{figure}

\begin{figure}[t!]
\includegraphics[angle=270, width=\linewidth]{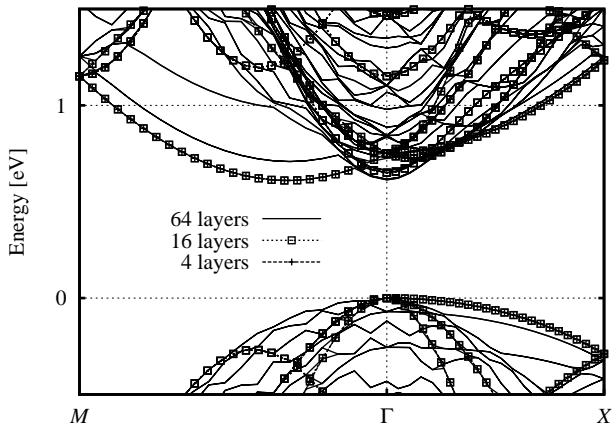}
\caption{Band structure diagram for the tetragonal bulk Si structures with increasing number of layers, calculated using the \textsc{vasp} plane wave method (see Sec. \ref{sec:methods}).}
\label{fig:d0vasp}
\end{figure}

Consider now the $\delta$-doping case discussed above (see Sec. \ref{sec:methods}), where we wish to align our cell with the $\left[1 1 0\right]$ and $\left[\bar{1} 1 0\right]$ directions (by rotating the cell 45$^{\circ}$ anticlockwise about $z$; this will also require a resizing of the cell in the plane to maintain periodicity -- see Fig. \ref{fig:rotation}), to allow us to include precisely four atoms per monolayer (as required for the minimal representation of 1/4 ML doping).  We now have a situation where the $X_{\rm{TET}}$ point in the new tetragonal BZ (see Fig. \ref{fig:tetBZ}) is no longer in the direction of the $X_{\rm{SC}}$ point in the simple cubic BZ, despite both $X$ points being in the centre of a face of their BZ.  Due to the rotation, what was the $\Delta_{\rm{SC}}$ direction in the simple cubic BZ is now the $\Sigma_{\rm{TET}}$ direction (pointing towards $M$, at the corner of the BZ in the $k_{z}=0$ plane) in the tetragonal BZ.  The tetragonal CBM, while physically still the same as the CBM in the FCC or simple cubic BZ, is not represented in the same fashion (see Fig. \ref{fig:d0vasp}).

\section{Band folding in $z$-direction}
\label{sec:zfold}

\begin{table}[b!]
  \begin{center}
    \begin{tabular}{ccccc}
      \hline
      \hline
      Basis            & No. of     & No. of            & LUMO                   & CBM       \\
      type             & layers     & \textit{k}-pts    & at $\Gamma$            & (at $\Delta_{\rm{FCC}}$)\\
                       &            & in $k_{z}$        & (eV)                   & (eV)      \\
      \hline
      PW               & 4          & 12                & 0.7517                 &           \\
      (\textsc{vasp})  & 8          & 6                 & 0.7517                 &           \\
                       & 16         & 3                 & 0.6506                 &           \\
                       & 32         & 2                 & 0.6170                 &           \\
                       & 40         & 1                 & 0.6179                 &           \\
                       & 64         & 1                 & 0.6137                 &           \\
                       & 80         & 1                 & 0.6107                 & 0.6102    \\
      \hline
      DZP              & 40         & 1                 & 0.6218                 &           \\
      (\textsc{siesta})& 60         & 1                 & 0.6194                 &           \\
                       & 80         & 1                 & 0.6154                 &           \\
                       & 120        & 1                 & 0.6145                 &           \\
                       & 160        & 1                 & 0.6151                 & 0.6145    \\
      \hline
      SZP              & 40         & 1                 & 0.8392                 &           \\
      (\textsc{siesta})& 60         & 1                 & 0.8349                 &           \\
                       & 80         & 1                 & 0.8315                 &           \\
                       & 120        & 1                 & 0.8311                 &           \\
                       & 160        & 1                 & 0.8315                 &           \\
                       & 200        & 1                 & 0.8310                 & 0.8309    \\
      \hline
      \hline
    \end{tabular}
  \end{center}
  \caption{Energy levels of tetragonal bulk Si structures.  (For details of calculation parameters, see Sec. \ref{sec:methods})}
  \label{tab:bulkSi}
\end{table}

Increasing the $z$-dimension of the cell leads to successive folding points being introduced as the Brillouin zone (BZ) shrinks along $k_{z}$ (see App. \ref{sec:bravais}).  This has the effect of shifting the conduction band minima in the $\pm k_{z}$ directions closer and closer to the $\Gamma$ point (see Fig. \ref{fig:folding}) and making the bandstructure extremely dense when plotting along $k_{z}$.  This results in the value of the lowest unoccupied eigenstate at $\Gamma$ being lowered as what were originally other sections of the band are successively mapped onto $\Gamma$, and after a sufficient number of folds the value at $\Gamma$ is indistinct from the original conduction band minimum (CBM) value. The effects of this can be seen in Table \ref{tab:bulkSi}, which describes increasingly elongated tetragonal cells of bulk Si. When we then plot the bandstructure in a different direction, \textit{e.g.} along $k_{x}$, the translation of the minima from $\pm k_{z}$ onto the $\Gamma$-point appear as a new band with two-fold degeneracy.  The degeneracy of the original band drops from 6- to 4-fold, in line with the reduced symmetry (we only explicitly calculate one, and the other three occur due to symmetry considerations).  This is the origin of the ``$\Gamma$-bands'' discussed in Refs. \onlinecite{Qian05} \& \onlinecite{Carter09}.  Once the $k_{z}$ valleys are sited at $\Gamma$, parabolic dispersion corresponding to the transverse kinetic energy terms is observed along $k_{x}$ and $k_{y}$, at least close to the band minimum (see Fig. \ref{fig:d0vasp}). 

All methods considered in Table \ref{tab:bulkSi} show the LUMO at $\Gamma$ (folded in along $\pm k_{z}$) approaching the CBM value as the amount of cladding increases; at 80 layers, the LUMO at $\Gamma$ is within 1 meV of the CBM value.  It is also of note that the PW indirect bandgap agrees well with the DZP value, and less so with the SZP model.  This is an indication that, although the behaviour of the LUMO with respect to the cell shape is well-replicated, the SZP basis set is demonstrably incomplete.  Conversely, pairwise comparisons between the PW and DZP results show agreement to within 5 meV.

It is important to distinguish effects indicating convergence with respect to cladding for doped cells (\textit{i.e.} elimination of layer-layer interactions) from those mentioned above which derive from the shape and size of the supercell.  Strictly, the convergence (with respect to the amount of encapsulating Si) of those results we wish to study in detail, such as the differences in energy between occupied levels in what was the bulk bandgap, provide the most appropriate measure of whether sufficient cladding has been applied.

\end{document}